# SYMMETRY AND MAGNETICALLY DRIVEN FERROELECTRICITY IN RARE-EARTH MANGANITES $RMnO_3$ (R=Gd, Tb, Dy)

J.L. Ribeiro

Centro de Física, Universidade do Minho, Campus de Gualtar, 4710-007 Braga, Portugal

This work investigates the magnetically driven ferroelectricity in orthorhombic manganites $RMnO_3$ (R=Gd, Dy or Tb) from the point of view of the symmetry. The method adopted generalizes the one used to characterize the polar properties of displacive modulated structures to the case of an irreducible magnetic order parameter. The symmetry conditions for magnetically induced ferroelectricity are established and the Landau-Devonshire free energy functionals derived from general symmetry considerations. The ferroelectric polarisation observed in $DyMnO_3$ and $TbMnO_3$ at zero magnetic field is explained in terms of the symmetry of a reducible magnetic order parameter. The polarisation rotation induced in these compounds by external magnetic fields and the stabilization of a ferroelectric phase in $GdMnO_3$ are accounted for by a mechanism in which magnetization and polarization are secondary order parameters that are not directly coupled but compete with each other through their coupling to competing primary modulated order parameters.

## I. Introduction

In the last decades, the search for novel materials displaying enhanced magneto-electric effect has attracted a great interest due to the potential technological application that can be envisaged and the subtlety of the physical mechanisms involved[1 2]. Until recently, the magneto-electric coupling in insulating and single phase materials has been explored mainly in the few known multiferroic compounds [3 4 5]. In materials of this type, such as $BiFeO_3$ [6 7] or $BiMnO_3$ [8], the co-existence of ferroelectric and magnetic cooperative orders in the same phase might prompt one to posit a relatively efficient coupling between the polar and magnetic degrees of freedom. However, because of the quite different critical temperatures associated with the magnetic and polar ordering, the experimentally observed magneto-electric coupling in these systems is weak and only very small signatures of the lower temperature magnetic phase are seen in the dielectric properties. In this respect, the behaviour observed may be considered similar to that found in anti-ferromagnetic and ferroelectric hexagonal manganites of the $YMnO_3$ family [9 10 11].

The situation has been dramatically altered with the discovery of magnetically induced ferroelectricity in several frustrated magnets, such as the perovskites $RMnO_3$ and $RMn_2O_5$ (R= rare earth element), $Ni_3V_2O_8$, delafossite $CuFeO_2$, spinel $CoCr_2O_4$, $MnWO_4$ and hexagonal ferrites $(Ba,Sr)_2Zn_2Fe_{12}O_{22}$ [12 13 14 15 16 17 18 19 20]. Here, in contrast to the conventional multiferroics, the paramagnetic phase is also paraelectric and the spontaneous ferroelectric polarization appears to be directly driven by a transition to a magnetic modulated phase. Not surprisingly, the observed magneto-electric coupling is much stronger and gives rise to a wealth of fascinating phenomena that are currently being investigated.

From a perspective based on symmetry and group theory, the idea of improper ferroelectricity driven by the condensation of a primary order parameter of a magnetic nature raises interesting questions. In the case of the novel frustrated magnets, a key point is to know under which circumstances a modulated magnetic order parameter can induce ferroelectricity. Improper ferroelectricity driven by displacive lattice modulation has been extensively investigated over two decades ago in molecular systems like $RbZn_2Cl_4$, $(C(NH_3)_4)_2CoCl_4$ or BCCD ($(CH_3)_3NCH_2COO.CaCl_2.2H_2O$). In these cases, it has been possible to take advantage of the simple crystallographic structure and irreducible nature of the modulated order parameter to obtain, from general symmetry considerations, common pictures of the structural phase transitions and precise predictions for the polar or non-polar character of a given modulated phase [21] [22] [23]. Particularly interesting is the case of BCCD, a compound that shares with the orthorhombic manganites $RMnO_3$ the same reference space group Pnma [a] and displays a wealth of structurally commensurate or incommensurate modulated phases. Here, a definite relationship between the modulation wavenumber and the direction of the electric polarization could be established [23].

A question naturally arises if the method of analysis used to investigate the polar properties of modulated displacive systems can be adapted to the case of the magnetoelectric compounds and to a situation where the primary order parameter is of a magnetic nature. The case of BCCD, for example, has been used to suggest a possible mechanism [24] for the magnetic field induced rotation of the electric polarization observed in TbMnO3 and DyMnO2 [24 25 26]. According to this suggestion, the electric polarization in these compounds would result from the secondary lattice modulation (of a displacive nature) magnetoelastically induced by the primary magnetic modulation. Consequently, a magnetic field could, by tuning the primary magnetic modulation, modify the secondary lattice modulation and, as a result, induce a rotation of the polarization. However, this key role of the lattice modulation is inconsistent with the well known fact that it is the primary order parameter that determines the symmetry of the ordered phase[27 28]. Hence, the displacive lattice modulation and the electric polarization must be both seen as secondary parameters whose direct coupling to the primary magnetic order parameter is restricted by general symmetry conditions.

---

[a] In this work, the standard setting Pnma is adopted for the RMnO3 compounds, as opposed to the Pbnm setting used in some articles.

In general, the complexity of the magnetic structures of most of the novel magneto-electric compounds makes the investigation of their magnetoelectric properties using symmetry considerations a difficult task. Among this set of materials, the rare earth manganites $RMnO_3$ (R=Gd, Tb or Dy) are relatively simple from a crystallographic point of view and their properties are well characterized experimentally. They therefore represent appropriate model systems to investigate further the mechanisms of the remarkable magneto-electric coupling observed in the class of frustrated magnets[20 24 25 26 29 30 31].

This work analyses the symmetry conditions that allow for the magnetically induced ferroelectricity observed in the orthorhombic rare-earth manganites $RMnO_3$ (R=Gd, Tb or Dy). From that analysis we derive adequate and symmetry based free energy functionals capable of describing both the ferroelectricity observed in $TbMnO_3$ and $DyMnO_3$ at zero magnetic field, the polarization rotation induced in these compounds by external magnetic fields and the ferroelectric phase induced in $GdMnO_3$ by an external magnetic field.

This paper is organized as follows. In section II we will briefly review the essential phenomenology of the magnetoelectric effect in the RMnO3 compounds. Section III analyses some general aspects related to the compatibility between a modulated magnetic order and a spontaneous polarization and establishes the methods of the analysis to be made. In section IV we will adapt the analysis made in [23] for BCCD to the case of a magnetic order parameter and use the complete irreducible co-representations of the magnetic space group of the reference phase to obtain the possible magnetic space groups of the commensurate phases that can be originated from the condensation of a magnetic irreducible order parameter. Sections V and VI analyse the possible translational invariants that can be constructed from the order parameter components and derive symmetry based Landau free energy functionals that are capable of describing the behaviour experimentally observed in the orthorhombic manganites.

## II. The magnetoelectric effect in the rare-earth manganites $RMnO_3$

The rare earth manganites $RMnO_3$ (R=Gd, Tb or Dy) are remarkable examples of the novel family of magneto-electric materials [12, 24, 32, 33, 34]. At room temperature, these compounds are paraelectric and paramagnetic with a distorted perovskite structure of orthorhombic symmetry (space group Pnma). All the three compounds undergo a phase transition to a magnetic incommensurate phase with a modulation wavevector directed along the a-axis ($\vec{k} = [\delta;0;0]$, $\delta(T_i) \approx 0.24$ (Gd), $\delta(T_i) \approx 0.28$ (Tb), $\delta(T_i) \approx 0.36$ (Dy); Ti $\approx 40K$) [32] [33] [34]. This incommensurate modulation corresponds to a collinear arrangement of the Mn magnetic momenta. With further cooling of the Tb and Dy compounds, under zero magnetic field, $\delta(T)$ approaches and enters into quasi-commensurate plateaus $(T \approx 20K)$ corresponding approximately to the values $\delta \approx 0.275$ and $\delta \approx 0.375$, for $TbMnO_3$ and $DyMnO_3$, respectively. The onset of these plateaus marks a transition from a collinear to a cycloidal modulation of the Mn magnetic momenta. This modulation remains directed along the same axis ($\vec{k} = [\delta;0;0]$) and gives rise to a ferroelectric polarization directed along the $\vec{b}$ crystallographic axis [12 24 35 36].

At zero magnetic field, $GdMnO_3$ remains paraelectric. The modulated magnetic order below $T_i$ remains collinear until the system enters into a A-type antiferromagnetic phase at about 27K. The application of a strong magnetic field ($H \geq 5$ Tesla) along the a-axis induces in this compound a ferroelectric phase with a polarization oriented along the c-axis. Structural studies indicate that this magnetically induced ferroelectric phase is accompanied by a commensurate modulation of the Mn momenta corresponding to $\vec{k} = \left[\frac{1}{4};0;0\right]$ [31]. Also, in the cases of $TbMnO_3$ and $DyMnO_3$, a similar magnetic field, applied along the a-axis, stabilizes a commensurate plateau with $\vec{k} = \left[\frac{1}{4};0;0\right]$ and induces a ferroelectric phase polar along the c-axis [12 24 34]. An external magnetic field is therefore capable of altering the symmetry of the magnetic structure and rotating the spontaneous polarization by 90º.

The $RMnO_3$ compounds also show, in addition to the magnetoelectric effects briefly described above, a strong lattice modulation originated from exchange striction. The modulate arrangement of the Mn spin moments with a wavevector $\vec{k} = [\delta;0;0]$ induces a lattice modulation with a wavevector $\vec{k}_{latt} = [2\delta;0;0]$ [12, 35]. The superlattice reflexions correspond therefore to a second harmonic of the magnetic modulation and, together with the electrical polarization, must be considered as secondary effects of a primary order parameter.

## III. Description of the method

To some extent, the elusive interaction between electric and magnetic degrees of freedom results from the different symmetries of the electric polarization (*P*) and magnetization (*M*) that are expressed by their different behaviour under spatial inversion (*i*) or time reversal ($\theta$). This different behaviour imposes, for example, that the coupling between static and homogeneous *P* and *M* must necessarily be non-linear ($\propto \pm P^{2n} M^{2m}$ , n and m integers). As a result, a homogeneous polarization can never result from the onset of a homogeneous magnetization because this would require the existence of coupling terms linear in *P* [28] that are, in this case, forbidden by symmetry. Therefore, if we restrict ourselves to time independent phenomena, a magnetically driven ferroelectricity may only eventually occur when the primary magnetic order parameter is spatially non-homogeneous.

Let us then consider the case of a modulated order parameter that is characterized by a single wavevector $\vec{k}$ located in the interior of the Brillouin zone and directed along a line of fixed symmetry (which corresponds to a $\Sigma$ or $\Lambda$ line, in the case of the $RMnO_3$ compounds or BCCD, respectively). This modulated order parameter can be written as:

$$\vec{S}(\vec{x}) = \vec{S}e^{i\vec{k}\cdot\vec{x}} + \vec{S}^{*}e^{-i\vec{k}\cdot\vec{x}} \qquad (1)$$

where, $\vec{S} = \vec{S}_0 e^{i\pi\Phi}$ and $\Phi$ is the phase defined with respect to the underlying discrete lattice. Note that, because $\vec{S}(\vec{x})$ is a real quantity, $\vec{S}^{*}_{\vec{k}} = \vec{S}_{-\vec{k}}$. Also, given its magnetic nature, $\theta\vec{S}(\vec{x}) = -\vec{S}(\vec{x})$. Under which conditions does this general modulated order parameter allow for mixed invariants linear in P and, therefore, for improper ferroelectricity?

A general nomial constructed from the complex conjugated components of $\vec{S}(\vec{x})$ has the form $(S)^{p}(S^{*})^{q}$. Naturally, if p=q, such a term is independent of the phase of the order parameter and gives rise to a homogeneous contribution to the free energy. Moreover, given its magnetic nature, it is necessarily even under inversion and does not allow for mixed invariants linear on P. We must therefore look for terms with $p \neq q$, that is, terms that depend on the phase $\Phi$ of the order parameter. Given that the polarization P is translational invariant, we must consider only those nomials that are themselves translationally invariants, in order to obtain possible coupling terms of the form $(S)^{p}(S^{*})^{q}P$. In general, these phase dependent terms can be written as (p and p´ integers):

$$f^{p}(S,S^{*}) = S^{p\acute{}}S^{*(p-p\acute{})} = S_o^{p}e^{i\pi(2p-p\acute{})} \qquad (2)$$

Consider now the requirement of translational invariance of (2). Let $\hat{T} = m\vec{a}_i$ be a Bravais translation of the reference phase, directed parallel to the modulation wavevector $\vec{k} = \delta\vec{a}_i^*$. Under such operation $f^p$ is transformed as:

$$\hat{T}f^p = f^p e^{-2\pi i\delta(p-2p')} \qquad (3)$$

Hence, for $f^p$ be invariant under this translation, $\delta(p-2p')$ must be an integer m. This is obviously not possible for $p \neq 2p'$ if $\vec{k}$ is incommensurate with respect to the reference lattice. An incommensurate order parameter does not allow for phase dependent translational invariants and, in consequence, cannot induce an electrical polarization. Incommensurate magnetic order and ferroelectricity are mutually incompatible.

On the other hand, if $\vec{k}$ is commensurate, the condition $\delta(p-2p') = m$ can be satisfied and non trivial translational invariants do exist. If, under the symmetry operations of the reference phase, any of these invariants is transformed as a polar vector, then a mixed invariant linear in $P$ can be constructed and improper ferroelectricity is allowed by symmetry. We can therefore verify if a given magnetic commensurate phase may or may not be polar by checking the transformation properties of the order parameter translational invariants.

Since we have established that improper ferroelectricity can only occur in the case of a commensurate phase, we may use an alternative method to investigate whether $P$ is a possible secondary order parameter. Given the symmetry of the order parameter by specifying the irreducible representation according to which it is transformed, we can directly calculate the possible magnetic space groups of the ordered phase and check if any of these groups is compatible with ferroelectricity. Let $G$ denote the unitary space group and $M = G \otimes \{E, \theta\}$ the magnetic space group of the paramagnetic reference phase, with $E$ and $\theta$ representing the identity and the time reversal operation, respectively. Let $\{g; \vec{t}\}$ and $\{\theta g; \vec{t}\}$ be elements of $G$ and $M - G$, respectively. By definition of order parameter, the symmetry of the ordered magnetic commensurate phase will be described by a magnetic space group M´ formed by the unitary and anti-unitary operators that leave the order parameter invariant up to a Bravais translation. That is, if $\hat{D}(g)$ and $\hat{D}(\theta g)$ are matrices that represent the operations $\{g; \vec{t}\}$ and $\{\theta g; \vec{t}\}$ in the (irreducible) space of the components of the order parameter and if the conditions

$$\hat{T} \times \hat{D}(g) \times \vec{S} = \vec{S} \tag{4a}$$

$$\hat{T} \times \hat{D}(\theta g) \times \vec{S}^* = \vec{S} \tag{4b}$$

are verified, then $\{g; \vec{t} + \vec{T}\}$ and $\{\theta g; \vec{t} + \vec{T}\}$ will belong to M´. The set of symmetry operations that verify (4a,b) and form the symmetry group of the ordered phase will, in general, depend on the symmetry of the order parameter, the commensurate value of $\delta$ (that is: on the odd (even) value of the integers in the fraction $\delta$) and the particular phase of the order parameter with respect to the underlying lattice.

We have therefore systematic methods for determining if a given commensurate magnetic order parameter can induce a ferroelectric polarization. They require the knowledge of the way the different elements of the symmetry group of the reference phase act on the linear space generated by the components of the order parameter. If we assume that the order parameter is irreducible, this amounts to knowing the complete irreducible co-representations (CICR) of the paramagnetic space group for a given commensurate vector $\vec{k}_c$ in the interior of the Brillouin zone. In the case pertaining to $RMnO_3$, the paramagnetic space group of the reference phase is (Pnma)´ and the modulation wavevector of the order parameter is $\vec{k} = (\delta(T),0,0)$. This vector corresponds to the wavevector $k_7$ in Kovalev`s tables [37]. Following the standard methods [37,38,39,40], we can readily obtain the CICR matrices that are given in Table 1 for the generators of the magnetic space group ($C_{2x}$, $\sigma_y$, $i$ and $i\theta$); Se also the appendix for the details of this calculation. The matrices corresponding to the other symmetry elements can then be obtained from the multiplication table of the group and by taking into account that the anti-unitary operators conjugate the coefficients of the matrices upon which they act.

**TABLE 1: Matrices representing the generators of the group (Pnma)´ in the four of its complete irreducible co-representations at $\vec{k} = \delta\vec{a}_1^*$.**

| | $C_{2x}$ | $\sigma_z$ | $i\theta$ | $i$ |
|---|---|---|---|---|
| $\Gamma(A_1)$ | $\begin{bmatrix} \varepsilon & 0 \\ 0 & \varepsilon^* \end{bmatrix}$ | $\begin{bmatrix} \varepsilon & 0 \\ 0 & \varepsilon^* \end{bmatrix}$ | $\begin{bmatrix} -1 & 0 \\ 0 & -1 \end{bmatrix}$ | $\begin{bmatrix} 0 & 1 \\ 1 & 0 \end{bmatrix}$ |
| $\Gamma(B_2)$ | $\begin{bmatrix} -\varepsilon & 0 \\ 0 & -\varepsilon^* \end{bmatrix}$ | $\begin{bmatrix} -\varepsilon & 0 \\ 0 & -\varepsilon^* \end{bmatrix}$ | $\begin{bmatrix} -1 & 0 \\ 0 & -1 \end{bmatrix}$ | $\begin{bmatrix} 0 & 1 \\ 1 & 0 \end{bmatrix}$ |
| $\Gamma(A_2)$ | $\begin{bmatrix} \varepsilon & 0 \\ 0 & \varepsilon^* \end{bmatrix}$ | $\begin{bmatrix} -\varepsilon & 0 \\ 0 & -\varepsilon^* \end{bmatrix}$ | $\begin{bmatrix} -1 & 0 \\ 0 & -1 \end{bmatrix}$ | $\begin{bmatrix} 0 & 1 \\ 1 & 0 \end{bmatrix}$ |
| $\Gamma(B_1)$ | $\begin{bmatrix} -\varepsilon & 0 \\ 0 & -\varepsilon^* \end{bmatrix}$ | $\begin{bmatrix} \varepsilon & 0 \\ 0 & \varepsilon^* \end{bmatrix}$ | $\begin{bmatrix} -1 & 0 \\ 0 & -1 \end{bmatrix}$ | $\begin{bmatrix} 0 & 1 \\ 1 & 0 \end{bmatrix}$ |

In the following sections we will apply these methods to characterize the potential polar properties of a magnetic commensurate phase and to deduce symmetry based free energy functionals that are adequate to describe the magnetoelectric effect in orthorhombic rare-earth manganites.

## IV. Possible magnetic space groups for the commensurate phases in RMnO3

As seen above, the magnetic space groups of the possible commensurate modulated phases, originating from the condensation of a given irreducible magnetic order parameter, can be directly obtained from the inspection of the set of unitary (*R*) and anti-unitary (*B*) symmetry operations that leave a given irreducible order parameter invariant up to a Bravais translation of the reference lattice. That is, if the equations:

$$\begin{bmatrix} e^{-2\pi jn\delta} & 0 \\ 0 & e^{2\pi jn\delta} \end{bmatrix} \times \begin{bmatrix} D_{11}(R) & D_{12}(R) \\ D_{21}(R) & D_{22}(R) \end{bmatrix} \times \begin{bmatrix} e^{j2\pi\Phi} \\ e^{-j2\pi\Phi} \end{bmatrix} = \begin{bmatrix} e^{j2\pi\Phi} \\ e^{-j2\pi\Phi} \end{bmatrix} \qquad (5a)$$

$$\begin{bmatrix} e^{-2\pi jn\delta} & 0 \\ 0 & e^{2\pi jn\delta} \end{bmatrix} \times \begin{bmatrix} D_{11}(B) & D_{12}(B) \\ D_{21}(B) & D_{22}(B) \end{bmatrix} \times \begin{bmatrix} e^{j2\pi\Phi} \\ e^{-j2\pi\Phi} \end{bmatrix}^{*} = \begin{bmatrix} e^{j2\pi\Phi} \\ e^{-j2\pi\Phi} \end{bmatrix} \qquad (5b),$$

are satisfied for a given *R* or *B*, then this element will belong to the magnetic group of the ordered phase. These groups will depend in general on the irreducible representation considered for the order parameter, on its phase and on the type of the modulation wavenumber, that is on the odd (even) value of the integers defining the fraction $\delta$. The results obtained for the case pertaining to the $RMnO_3$ compounds are listed in the Table 2. Here, the magnetic space groups are denoted as *M(G)* if $\theta$ is not a symmetry element or $(G)' = G \otimes \{E, \theta\}$ if $\theta$ is a symmetry element of the group.

It is interesting to consider the particular case of time reversal and spatial inversion because these operations directly provide us with two necessary conditions for the existence of ferroelectricity. Moreover, in the particular case under analysis, the matrices corresponding to these two operations are found to be independent of the particular irreducible co-representation considered for the order parameter. Let us see first the case of time reversal. From the matrix representing this operation one finds from (5b) that $\theta$ is a symmetry of the ordered phase if and only if $\delta_c = \frac{odd}{even}$. Therefore, in the case under analysis, only this type of irreducible magnetic order parameter can induce improper ferroelectricity.

For spatial inversion one finds that, independently of the modulation wavenumber type, the operation is always a symmetry element of the commensurate phase if $\Phi = 0$. In this case, a ferroelectric polarization is obviously not allowed but an improper ferromagnetic magnetization $M_i$ may possibly occur. In fact, for $\delta = \frac{odd}{odd}$, $M_x$, $M_y$ or $M_z$ are allowed if $\Phi = 0$ and the irreducible order parameter has a symmetry $\Gamma(B_1)$, $\Gamma(A_1)$ or $\Gamma(A_2)$, respectively., while for $\delta = \frac{even}{odd}$, $M_x$, $M_y$ or $M_z$ are allowed if $\Phi = 0$ and the irreducible order parameter has a symmetry $\Gamma(A_2)$, $\Gamma(B_1)$ or $\Gamma(A_1)$. For $\delta = \frac{odd}{even}$ no homogeneous magnetization is allowed because of the existence of time reversal symmetry. From these results it may be possible to identify the symmetry of the irreducible order parameter related to a given commensurate phase from the knowledge of the modulation wavenumber and magnetization of that phase.

**TABLE 2: The possible magnetic space groups of the commensurate phases for the different symmetries of the order parameter.**

| $\delta = \frac{2k+1}{2m+1}$ | $\Phi = 0$ | $\Phi = \frac{\pi}{2}$ |
|---|---|---|
| $\Gamma(A_1)$ | $Pnma(P1\frac{2_1}{m}1)$ | $Pnma(Pnm2_1)$ |
| $\Gamma(B_2)$ | $Pnma(Pnma)$ | $Pnma(P2_1ma)$ |
| $\Gamma(A_2)$ | $Pnma(P11\frac{2_1}{a})$ | $Pnma(Pn2_1a)$ |
| $\Gamma(B_1)$ | $Pnma(P\frac{2_1}{n}11)$ | $Pnma(P2_12_12_1)$ |
| $\delta = \frac{2k}{2m+1}$ | $\Phi = 0$ | $\Phi = \frac{\pi}{2}$ |
| $\Gamma(A_1)$ | $Pnma(Pnma)$ | $Pnma(P2_1ma)$ |
| $\Gamma(B_2)$ | $Pnma(P1\frac{2_1}{m}1)$ | $Pnma(Pnm2_1)$ |
| $\Gamma(A_2)$ | $Pnma(P\frac{2_1}{n}11)$ | $Pnma(P2_12_12_1)$ |
| $\Gamma(B_1)$ | $Pnma(P11\frac{2_1}{a})$ | $Pnma(Pn2_1a)$ |
| $\delta = \frac{2k+1}{2m}$ | $\Phi = 0$ | $\Phi = \frac{2k+1}{4m}\pi$ |
| $\Gamma(A_1)$ | $(P1\frac{2_1}{a}1)'$ | $(Pna2_1)'$ |
| $\Gamma(B_2)$ | $(P1\frac{2_1}{a}1)'$ | $(Pna2_1)'$ |
| $\Gamma(A_2)$ | $(P1\frac{2_1}{a}1)'$ | $(Pna2_1)'$ |
| $\Gamma(B_1)$ | $(P1\frac{2_1}{a}1)'$ | $(Pna2_1)'$ |

Also, in the case of $\delta = \frac{odd}{odd}$ or $\delta = \frac{even}{odd}$, time reversal symmetry is broken and ferroelectricity is not allowed even if the unitary sub-group is compatible with the onset of a spontaneous polarization. In fact, as we will see in the next section, the symmetry of the order parameter does not allow, in these cases, coupling terms linear in $P$. Interestingly, the symmetry of a commensurate phase with $\delta = \frac{2k+1}{2m}$ is independent of the irreducible representation of the order parameter and allows for a ferroelectric polarization directed along c-axis if $\Phi = \frac{2k+1}{4m}\pi$.

## V. Landau free energy density and magnetoelectric coupling for an irreducible order parameter: the case of GdMnO3

The potential polar properties of a magnetic commensurate phase induced by a given irreducible order parameter can also be investigated by analysing the transformation properties of the translational invariants. This alternative method is adequate if one wants to discuss the thermodynamics of the phase transition within the scope of the Landau-Devonshire theory. In this section we will briefly discuss this approach in order to characterize the type of magnetoelectric coupling that may be at the origin of a ferroelectricity induced by a magnetic irreducible order parameter.

As seen in section-III, for a commensurate phase with $\delta = \frac{n}{m}$, the non-trivial translational invariants that depend on the phase of the order parameter are nomials of the form $(S)^p (S^*)^q$ ( $p \neq q$ ). In this case, given that $S$ and $S^*$ are complex numbers, the analytical form of these translational invariants is determined by the image induced in the complex plane by the translational sub-group of the reference phase. For a wavenumber $\delta = \frac{n}{m}$, this image is isomorphic to the group *Cm* [41]. Then, as shown in [27 41], any homogeneous polynomial defined in the space of the complex numbers that is invariant under *Cm* may be expressed as linear combinations of terms of the type $S_0^m \cos(m\Phi)$ and $S_0^m \sin(m\Phi)$. Because of the different parity of these terms under spatial inversion, a coupling term linear in P must necessarily be of the form $PS_0^m \sin(m\Phi)$. Consider now the restrictions imposed by the requirement that the mixed term be invariant under time reversal. Because $\theta P = P$ and $\theta S_0 = -S_0$, the mixed term will be invariant only if *m* is even. That is, in agreement with the analysis of the preceding section, only magnetic commensurate phases of the type $\delta = \frac{odd}{even}$ can support a ferroelectric polarisation. Conversely, the translational invariants of the form $S_0^m \cos(m\Phi)$, being even under spatial inversion and odd under time reversal if *m* is odd, allow for mixed invariants linear on a magnetization. In this case improper ferromagnetism may occur.

With the inclusion of terms of the form $S_0^{2m} \cos(2m\Phi)$, which are globally invariant and give rise to *Umklapp* terms in the free energy density whenever $\delta$ is commensurate, the trivial invariants of the type $S_0^{2n}$ and additional magnetic or electric terms, one finds the following Landau free energy densities :

$$f_1 = \frac{\alpha}{2} S_0^2 + \frac{\beta}{4} S_0^4 + \cdots + \xi M S_0^m \cos(m\Phi) + \gamma S_0^{2m} \cos(2m\Phi) + \Omega_M S_0^2 M^2 + \frac{M^2}{2\chi_M} - MH \quad (6)$$

if $\delta \neq \frac{odd}{even}$ and :

$$f_2 = \frac{\alpha'}{2} S'^2_0 + \frac{\beta'}{4} S'^4_0 + \cdots + \gamma' S_0^{2m} \cos(2m\Phi) + \Omega'_M S'^2_0 M^2 + \frac{M^2}{2\chi_M} - MH + \cdots$$
$$+ \upsilon P S'^m_0 \sin(m\Phi) + \Omega_P S'^2_0 P^2 + \frac{P^2}{2\chi_P} \quad (7)$$

if $\delta = \frac{odd}{even}$.

In order to discuss the thermodynamics of the lock-in transition, we need to include the analysis of the possible stability of an incommensurate phase (although these phases are necessarily non-polar). For this purpose, the incommensurate phase can be described as a distorted commensurate phase. Consequently, we must include in the free energy expansion terms that depend on the spatial derivatives of the order parameter. If we assume that the spatial variations of the order parameter are slow, we can consider only the contributions of its first order derivatives and limit ourselves to lower order terms. In such a case we find the following four possibilities: $(\frac{\partial \vec{S}}{\partial x}); (\frac{\partial \vec{S}^*}{\partial x}); (\vec{S}\frac{\partial \vec{S}^*}{\partial x} \pm \vec{S}^* \frac{\partial \vec{S}}{\partial x})$ and $(\frac{\partial \vec{S}}{\partial x}) \times (\frac{\partial \vec{S}^*}{\partial x})$. The first two of these terms are naturally forbidden by symmetry because are not invariant under spatial inversion, a symmetry operation of the space group of the paramagnetic phase. The third term with the sign "+" can be written as $\frac{1}{2}\frac{d(\vec{S}\cdot\vec{S}^*)}{dx}$ and gives rise to a contribution to the free energy that is in fact homogeneous ($\Delta F \propto \int \frac{d(\vec{S}\cdot\vec{S}^*)}{dx}dx \propto S_0^2$). The fourth term $(\frac{\partial\vec{S}}{\partial x})\times(\frac{\partial\vec{S}^*}{\partial x}) = \left[S_0^2(\frac{\partial\Phi}{\partial x})^2 + (\frac{\partial S_0}{\partial x})^2\right]$ does not depend on the phase of the order parameter and is always allowed by symmetry. In particular, it is invariant under time reversal and spatial inversion. The third term with the sign "-", the Lifshitz term, may or may not be allowed depending on the symmetry of the system. In the case of the rare-earth manganites $RMnO_3$, because the prototype space group contains inversion and the wavevector $\vec{k}$ space is located inside the Brillouin zone, such an invariant is allowed [27]. In fact, under the spatial symmetry operations of the space group of the reference phase, $\vec{S}\partial\vec{S}^* - \vec{S}^*\partial\vec{S}$ transforms exactly as $\vec{x}$ if $\vec{k} // \vec{x}$. However, under time reversal, $\vec{S}\partial\vec{S}^* - \vec{S}^*\partial\vec{S}$ changes sign. Therefore, the Lifshitz invariant must be explicitly written as $j(\vec{S}\frac{\partial\vec{S}^*}{\partial x} - \vec{S}^*\frac{\partial\vec{S}}{\partial x}) = S_0^2(\frac{\partial\Phi}{\partial x})$, where $j = \sqrt{-1}$ ([b]).

With these terms allowing for incommensurate distortions of the commensurate phases, the two types of free energy densities are:

---

[b] Obviously, the two relevant terms [ $(\frac{\partial\vec{S}}{\partial x})\times(\frac{\partial\vec{S}^*}{\partial x})$ and $j(\vec{S}\frac{\partial\vec{S}^*}{\partial x} \pm \vec{S}^*\frac{\partial\vec{S}}{\partial x})$ ] are invariant under the unitary or non-unitary symmetry operations of the magnetic space group of the reference phase and cannot be linearly coupled to the electric polarization. For example, a term of the type $P(S\frac{\partial S^*}{\partial x} - S^*\frac{\partial S}{\partial x})$ is not possible if *P* is not itself invariant under all the spatial operations of the reference group.

$$f_1 = \frac{\alpha}{2}S_0^2 + \frac{\beta}{4}S_0^4 - \eta S_0^2(\frac{\partial\Phi}{\partial x}) + \frac{k}{2}\left(S_0^2(\frac{\partial\Phi}{\partial x})^2 + \left(\frac{\partial S_0}{\partial x}\right)^2\right) + \\ + \xi M S_0^m \cos(m\Phi) + \gamma S_0^{2m}\cos(2m\Phi) + \Omega_M S_0^2 M^2 + \frac{M^2}{2\chi_M} - MH \tag{8}$$

$$f_2 = \frac{\alpha'}{2}S'^2_0 + \frac{\beta'}{4}S'^4_0 + -\eta S'^2_0(\frac{\partial\Phi}{\partial x}) + \frac{k}{2}\left(S'^2_0(\frac{\partial\Phi'}{\partial x})^2 + \left(\frac{\partial S'_0}{\partial x}\right)^2\right) + \gamma' S'^{2m}_0\cos(2m\Phi') + \\ + \Omega'_M S'^2_0 M^2 + \frac{M^2}{2\chi'_M} - MH + \cdots + \upsilon P S'^m_0 \sin(m\Phi') + \Omega_P S'^2_0 P^2 + \frac{P^2}{2\chi_P} \tag{9),}$$

for $\delta \neq \frac{odd}{even}$ and $\delta = \frac{odd}{even}$, respectively.

Let us now see the case of $GdMnO_3$. The magnetic modulation observed in this compound below $T_i$ corresponds to a simple collinear modulation of the magnetic momenta of the Mn ions. If we assume that this structure results from the condensation of a irreducible magnetic order parameter, then the previous results show that a magnetically driven ferroelectricity is only possible in the case of a magnetic modulated commensurate phase of the type $\delta = \frac{odd}{even}$. Consistently, the polar phase induced in this compound by an external field corresponds precisely to the stabilization of the lowest possible order of this type of modulation ($\delta = \frac{1}{4}\vec{a}_1^*$). However, at zero field, the system remains paraelectric. Let us now analyse how the free energy densities given above may allow us to describe the stabilization of the ferroelectric phase by an external magnetic field.

In equations (8) and (9), the magnetization $M$ and the polarization $P$ are secondary order parameters that can be eliminated by imposing the conditions $\frac{\partial f_1}{\partial M}=0$ and $\frac{\partial f_2}{\partial M}=\frac{\partial f_2}{\partial P}=0$, corresponding to thermal equilibrium. These equations lead to:

$$\frac{\partial f_1}{\partial M}=0 \Rightarrow M=\overline{\chi}_M\left[H-\xi S_0^m\cos(m\Phi)\right]$$

$$\frac{\partial f_2}{\partial M}=0 \Rightarrow M=\overline{\chi}'_M\,H \tag{10}$$

$$\frac{\partial f_2}{\partial P}=0 \Rightarrow P=-\nu\overline{\chi}_P S'^m_0\sin(m\Phi'),$$

where, $\overline{\chi}_M=\frac{\chi_M}{1+2\Omega_M\chi S_0^2}$, $\overline{\chi}'_M=\frac{\chi'_M}{1+2\Omega'_M\chi S'^2_0}$ and $\overline{\chi}_P=\frac{\chi_P}{1+2\Omega_P\chi_P S'^2_0}$ represent renormalized magnetic and electric susceptibilities. The substitution of (10) into (8-9) gives:

$$\begin{aligned} f_1 &= \frac{\alpha}{2}S_0^2+\frac{\beta}{4}S_0^4-\eta S_0^2(\frac{\partial\Phi}{\partial x})+\frac{k}{2}\left(S_0^2(\frac{\partial\Phi}{\partial x})^2+\left(\frac{\partial S_0}{\partial x}\right)^2\right)+\\ &+(\gamma-\frac{\xi^2\overline{\chi}_M}{2})S_0^{2m}\cos^2(m\Phi)-\gamma\sin^2(m\Phi)-\frac{1}{2}\overline{\chi}_M H^2+\xi S_0^m\overline{\chi}_M H\cos(m\Phi)\end{aligned} \tag{11}$$

$$\begin{aligned} f_2 &= \frac{\alpha'}{2}S'^2_0+\frac{\beta'}{4}S'^4_0+-\eta S'^2_0(\frac{\partial\Phi'}{\partial x})+\frac{k}{2}\left(S_0^2(\frac{\partial\Phi'}{\partial x})^2+\left(\frac{\partial S'_0}{\partial x}\right)^2\right)+\gamma' S'^{2m}_0\cos^2(m\Phi')-\\ &-(\gamma'+\frac{\overline{\chi}_P\nu^2}{2})S'^{2m}_0\sin^2(m\Phi')-\frac{1}{2}\overline{\chi}'_M H^2\end{aligned} \tag{12}$$

In the case of $f_1$, which expresses the free energy density for a commensurate phase with $\delta\neq\frac{odd}{even}$, one can directly see that the phase of the order parameter $\Phi=0$ is favoured if $\gamma < 0$. In such a case, a spontaneous magnetization proportional to $S_0^m$ may occur. In addition, the application of an external magnetic field gives rise to a magnetostatic term that is linear in *H*. The effect of such a term in the relative stability of the phase is determined by the sign of the coefficient $\xi$. If $\xi$ is positive, the field increases the energy and tends to destabilize the phase. On another hand, in the case of a commensurate phase $\delta = \frac{odd}{even}$, such a term is forbidden by symmetry. Here, the coupling between the external magnetic field and the order parameter is simply expressed in $f_2$ by a renormalization of the magnetic susceptibility and, consequently, only the usual magnetostatic term $\propto H^2$ is allowed.

The above functionals suggest therefore a simple mechanism for the stabilization of a ferroelectric phase by an external magnetic field in the case of $GdMnO_3$. Consider for example a set of modulated phases of a common symmetry except for the different wavelength of the magnetic modulation. Below a certain temperature, several modulations may correspond to relative minima of the free energy and compete for absolute stability. If, at a given temperature and zero magnetic field, the free energy density of the stable commensurate phase corresponds to (6) with $\xi > 0$, then an external field increases the energy of the phase and may give rise to a first order transition to another commensurate phase with a different modulation wavelength, provided this other phase is less affected by the field. As seen, this can only occur if the modulation wave is of the type $\delta = \frac{odd}{even}$, for which a linear coupling in the field is not allowed by symmetry. In other words, above a certain threshold, the cost in energy required to change the wavelength of the modulation wave to a value compatible with ferroelectricity is smaller than that necessary to maintain the low field stable magnetic modulation. Magnetization and polarization are secondary order parameters that are not directly coupled but compete with each other via their coupling with competing primary modulated order parameters.

## VI. The case of a reducible order parameter: $TbMnO_3$ and $DyMnO_3$

The analysis given above can also be applied to the case of the similar systems $(Tb,Dy)MnO_3$. Let us consider first the situation at zero magnetic field in which the ferroelectric phase, observed in these compounds below $T_c \approx 28K$, corresponds to a cycloidal magnetic structure. For $28K < T < 41K$ there exist an incommensurate collinear modulation that is non-polar, as expected given the incommensurate nature of the magnetic order.

The cycloidal ferroelectric structure can be seen as resulting from the simultaneous condensation of two irreducible order parameters $S_i$ and $S_j$ of symmetry $\Gamma_i$ and $\Gamma_j$, respectively. This phase is often referred to as incommensurate [25,42] because $\delta(T)$ shows a quasi-plateau and slightly depends on temperature [25]. However, given its polar nature, this phase may be rather seen as a disordered commensurate phase in which the C-domains are separated by discommensurations, that is, regions where the phase of the order parameter varies rapidly. In this case, the spontaneous polarisation is originated from the commensurate regions that are characterized by a reducible order parameter that is transformed as the Krönecker product of its components ($\Gamma = \Gamma_i \otimes \Gamma_j$). Then, the onset of an electric polarisation along a given crystallographic direction ($\vec{P}_k$) is allowed only if $\Gamma_i \otimes \Gamma_j \otimes \Gamma^*(\vec{P}_k)$ contains the totally symmetric representation of the space group [43 44 45 46]. As shown in [43], this requires, as a necessary condition, that the sum over the wavevectors characterizing the three order parameters vanishes ($\vec{k}_i + \vec{k}_j + \vec{k}_{P_k} = 0$). Given that $\vec{k}_{P_k} = 0$, this selection rule imposes that we must consider all pairs of vectors of the stars of $\vec{k}_i$ and $\vec{k}_j$ such that $\vec{k}_i + \vec{k}_j = 0$. In terms of the transformation properties of possible mixed invariants involving the components of $S_i$, $S_j$ and $P_k$, this condition imposes that the lowest order coupling between the reducible order parameter and an electric polarisation must necessarily involve combinations of mixed forms of the type $S_i S_j^* P_k$ and $S_i S_j^* P_k$. We can therefore investigate the possible onset of a ferroelectric polarization by analysing if the translational invariants of the type $S_i S_j^* \pm S_j S_i^*$ transforms as components of a polar vector. This can be easily done with the help of the matrices of Table 1 and the results are given in Table 3. Note that the transformation properties of these invariants are similar to those of the symmetric (+) or anti-symmetric (-) Dzyaloshinskii-Moriya interaction $\vec{D} \cdot \vec{S}_n \times \vec{S}_{n+1}$ [47, 48]. In particular, they are always even under time reversal because correspond to the product of two components that are odd under this operation. Therefore, we need only to consider in Table 3 the unitary symmetry operations of the paramagnetic space group.

As seen in Table 3b, the translational invariants of the form $S_i S_j^* + S_j S_i^*$ are always even under inversion. They give rise to a coupling of the type $(S_i S_j^* + S_j S_i^*)\eta_{kl}(\vec{q} = 0)$ between the magnetic order parameter and homogeneous lattice strains $\eta_{kl}(\vec{q} = 0)$, the symmetry of which depends on the symmetry of the primary order parameter. Note that the transformation properties of terms of the type $S_i S_j + S_j^* S_i^*$ , to which corresponds a wavenumber $|\vec{q}| = |2\vec{k}|$, are similar to those given in Table 3b. These terms are of the form $\left[S_i S_j \eta_{kl}(-2\vec{k}) + S_j^* S_i^* \eta_{kl}(2\vec{k})\right]$ and are responsible for the onset of a lattice modulation with one half of the wavelength of the magnetic modulation. The symmetry of the primary order parameter determines the nature of the secondary modulated lattice strain.

**TABLE 3: Transformation properties of the lowest order mixed translational invariants in the case of a reducible order parameter of symmetry $\Gamma = \Gamma_i \otimes \Gamma_j$**

| | | $\Gamma_1$ | $\Gamma_2$ | $\Gamma_3$ | $\Gamma_4$ |
|---|---|---|---|---|---|
| | $\Gamma_1$ | – | $B_{1u}$ | $A_u$ | $B_{2u}$ |
| a) $S_i S_j^* - S_j S_i^*$ | $\Gamma_2$ | $B_{1u}$ | – | $B_{2u}$ | $A_u$ |
| | $\Gamma_3$ | $A_u$ | $B_{2u}$ | – | $B_{1u}$ |
| | $\Gamma_4$ | $B_{2u}$ | $A_u$ | $B_{1u}$ | – |
| | | $\Gamma_1$ | $\Gamma_2$ | $\Gamma_3$ | $\Gamma_4$ |
| | $\Gamma_1$ | $A_g$ | $B_{2g}$ | $B_{3g}$ | $B_{1g}$ |
| b) $S_i S_j^* + S_j S_i^*$ | $\Gamma_2$ | $B_{2g}$ | $A_g$ | $B_{1g}$ | $B_{3g}$ |
| | $\Gamma_3$ | $B_{3g}$ | $B_{1g}$ | $A_g$ | $B_{2g}$ |
| | $\Gamma_4$ | $B_{1g}$ | $B_{3g}$ | $B_{2g}$ | $A_g$ |

For example, for an order parameter of symmetry $\Gamma_2 \otimes \Gamma_3$ the modulated strain corresponds to the component $\eta_{xy}$ of the strain tensor (symmetry $B_{1g}$). This shear strain reflects the breaking of the orthorhombic symmetry that necessarily occurs in the case of a reducible order parameter. Also, in the case of a single irreducible order parameter, terms of the type $\left[S_i S_i \eta_{jj}(-2\vec{k}) + S_i^* S_i^* \eta_{jj}(2\vec{k})\right]$ are allowed. Because $S_i S_i + S_i^* S_i^*$ is transformed according to the totally symmetric representation Ag of the group $D_{2h}$, only the diagonal elements of the strain tensor are, in this case, possible secondary order parameters.

On another hand, the mixed translational invariants $S_i S_j^* - S_j S_i^*$ are odd under inversion and, with the exceptions of the symmetries $\Gamma_1 \otimes \Gamma_3$ and $\Gamma_2 \otimes \Gamma_4$, are transformed as components of a polar vector (symmetry $B_{1u}$ or $B_{2u}$). This means that polarizations oriented along the b- or the c-axis, depending on the symmetry of the magnetic order parameter, are allowed by symmetry but a polarization oriented along the a-axis (that is, parallel to the modulation wavevector) is forbidden. The particular case of an induced polarisation along the b-axis (symmetry $B_{2u}$), as it is experimentally observed in TbMnO3 and DyMnO3 at zero field, can be obtained for a primary order parameter with a symmetry $\Gamma_2 \otimes \Gamma_3$ or $\Gamma_1 \otimes \Gamma_4$. Note that neutron diffraction data [25] suggest that is the first possibility ($\Gamma_2 \otimes \Gamma_3$) that occurs in TbMnO3. If, for example, we assume that this is the case, we can immediately write the Landau free energy density corresponding to a given modulation wavenumber.

Let us consider, as one illustrative example, the magnetoelectric effects in the case of homogeneous commensurate phases (that is, assume that $\frac{\partial S_i}{\partial x} = \frac{\partial \Phi_i}{\partial x} = 0$) for a magnetic order parameter of symmetry $\Gamma_2 \otimes \Gamma_3$ . In such a case, we must consider the magnetization and the electrical polarisation as the relevant secondary order parameters. The other possible secondary effects, such as the magnetoelastic effects and the magnetically induced lattice modulation, can be ignored. Accordingly, and for simplicity sake, we will not consider the magnetoelastic terms of the type $(S_i S_j^* + S_j S_i^*)\eta_{kl}(\vec{q}=0)$ or $\left[S_i S_j \eta_{kl}(-2\vec{k}) + S_j^* S_i^* \eta_{kl}(2\vec{k})\right]$ in the free energy functionals. As seen, we must analyse separately the different types of modulation wavenumber $\delta$ . We can therefore write for the cases of $\delta = \frac{2k+1}{2m+1}$ or $\delta = \frac{2k}{2m+1}$ the free energy densities $f_3$ and $f_4$ given by equations 13a) and 13b), respectively:

$$f_3 = \frac{\alpha}{2}S_2^2 + \frac{\beta}{4}S_2^4 + ... + \gamma S_2^{2m}\cos(2m\Phi_2) + ... + \frac{\alpha'}{2}S_3^2 + \frac{\beta'}{4}S_3^4 + \gamma' S_3^{2m}\cos(2m\Phi_3) + \\ + M^2\left(\Omega_M S_2^2 + \Omega'_M S_3^2\right) + \sum_i \frac{M_i^2}{2\chi_{ii\,M}} - \vec{M}\cdot\vec{H} + \xi M_z S_3^m \cos(m\Phi_3) + \\ + \sigma P_y (S_2 S_3^* - S_3 S_2^*) + P_y^2(\Omega_P S_2^2 + \Omega'_P S_3^2) + \frac{P_y^2}{2\chi_P} \tag{13a}$$

$$f_4 = \frac{\alpha}{2}S_2^2 + \frac{\beta}{4}S_2^4 + ... + \gamma S_2^{2m}\cos(2m\Phi_2) + ... + \frac{\alpha'}{2}S_3^2 + \frac{\beta'}{4}S_3^4 + \gamma' S_3^{2m}\cos(2m\Phi_3) + \\ + M^2\left(\Omega_M S_2^2 + \Omega'_M S_3^2\right) + \sum_i \frac{M_i^2}{2\chi_{ii\,M}} - \vec{M}\cdot\vec{H} + \xi M_y S_2^m \cos(m\Phi_2) + \\ + \xi' M_x S_3^m \cos(m\Phi_3) + + \sigma P_y (S_2 S_3^* - S_3 S_2^*) + \left(\Omega_P S_2^2 + \Omega'_P S_3^2\right)P_y^2 + \frac{P_y^2}{2\chi_P} \tag{13b}$$

Here, we have taken into account that, for the considered symmetry of the order parameter, only the magnetization $M_z$ is allowed for $\delta = \frac{2k+1}{2m+1}$ while $M_x$ and $M_y$ can occur if $\delta = \frac{2k}{2m+1}$. Also, because $S_i S_j^* - S_j S_i^*$ is independent of $\vec{k}$, $P_y$ is in both cases a possible secondary order parameter, whose value can be obtained from the condition $\frac{\partial f}{\partial P_y} = 0$:

$$P_y = \frac{-\sigma\left(S_2 S_3^* - S_3 S_2^*\right)}{2\left(\Omega_P S_2^2 + \Omega'_P S_3^2\right) + \chi_P^{-1}} \qquad (14),$$

As seen, if $S_2 = 0$ or $S_3 = 0$, $P_y = 0$, that is, this polarisation vanishes in the case of an irreducible order parameter. Also, $P_y = 0$ if the two order parameters $S_2$ and $S_3$ are in phase ($\Phi_2 - \Phi_3 = 0$) and $P_y$ will be maximized if the two order parameters have a phase difference of $\Phi_2 - \Phi_3 = \frac{\pi}{2}$

For the reasons clarified in the preceding sections, the case $\delta = \frac{2k+1}{2m}$ must be considered separately. Here, on one hand, a linear coupling of the primary order parameters with a magnetization is forbidden by symmetry. On another hand, and as seen in section IV, we have always $P_z$ as a secondary order parameter. Hence, because $S_i S_j^* - S_j S_i^*$ does not depend on the wavenumber, we have here a potential competition between two polarisations $P_y$ and $P_z$. For $\Gamma = \Gamma_2 \otimes \Gamma_3$ and $\delta = \frac{2k+1}{2m}$, the free energy density must be written as:

$$\begin{aligned} f_4 &= \frac{\alpha}{2} S_2^2 + \frac{\beta}{4} S_2^4 + \ldots + \gamma S_2^{2m} \cos(2m\Phi_2) + \ldots + \frac{\alpha'}{2} S_3^2 + \frac{\beta'}{4} S_3^4 + \gamma' S_3^{2m} \cos(2m\Phi_3) + \\ &+ M^2\left(\Omega_M S_2^2 + \Omega'_M S_3^2\right) + \sum_i \frac{M_i^2}{2\chi_{ii\,M}} - \vec{M} \cdot \vec{H} + \sigma P_y (S_2 S_3^* - S_3 S_2^*) + \\ & P_z\left[\upsilon S_2^m \sin(m\Phi_2) + \upsilon' S_3^m \sin(m\Phi_3)\right] + (\Omega_P S_2^2 + \Omega'_P S_3^2)(P_y^2 + P_z^2) + \frac{P_y^2}{2\chi_{yy\,P}} + \frac{P_z^2}{2\chi_{zz\,P}} \end{aligned} \qquad (15)$$

If $\gamma, \gamma' > 0$, the phases $\Phi_2 = \Phi_3 = \frac{\pi}{2}$ will be favoured. In this case, $P_y = 0$ and

$$P_z = \frac{-\left[\upsilon S_2^m \sin(m\Phi_2) + \upsilon' S_3^m \sin(m\Phi_3)\right]}{2(\Omega_P S_2^2 + \Omega'_P S_3^2) + \frac{1}{\chi_{zzP}}}.$$

Let us finally consider the effect of an external magnetic field and the rotation of the polarization from $P_y$ to $P_z$ that is experimentally observed in TbMnO3. The mechanism suggested by (13-15) for the rotation of the polarization under a magnetic field is entirely similar to that described for the stabilization of the ferroelectric phase in GdMnO3. Here again, due to the presence of coupling terms linear on $M_z$ if $\delta = \frac{2k+1}{2m+1}$ or on $M_x$ and $M_y$ if $\delta = \frac{2k}{2m+1}$, the magnetostatic energy of the commensurate phase stable at zero field, polar along the b-axis, can be strongly increased by an external field applied along the c-axis (if $\delta = \frac{2k+1}{2m+1}$) or along the a- or b-axes (if $\delta = \frac{2k}{2m+1}$), if $\xi > 0$ or $\xi' > 0$ . Consequently, above a certain threshold, it may become energetically favourable to diminish this magnetostatic energy by slightly adjusting the value of modulation wavenumber and the phase of $S_2$ and $S_3$ to the values ($\delta = \frac{2k+1}{2m}$ ) and $\Phi_2 = \Phi_3 = \frac{\pi}{2}$, respectively, for which terms linear on $M$ are forbidden. As a consequence of this discontinuous transition the polarisation rotates from the b-axis to the c-axis.

## VII- CONCLUSION

The analysis made in the present work of the magnetoelectric coupling in the orthorhombic manganites RMnO3 is purely phenomenological and based on general symmetry arguments. The method adopted allows us to draw several general conclusions about the compatibility between magnetic modulated order and ferroelectricity. We have seen that the lattice modulation observed in the rare-earth compounds is not an essential ingredient for the stabilization of the ferroelectric state. Being a secondary effect it cannot determine the symmetry or the polar properties of the ordered phase. Note that although an incommensurate magnetic modulation, for example, may induce a lattice modulation with $\delta_{latt.} = 2\delta$, it is incompatible with the onset of a spontaneous polarisation. This incompatibility is solely determined by the primary order parameter: symmetry forbids in this case a linear coupling between the order parameter and *P*. We have also shown that from the vantage point of symmetry, an irreducible commensurate order parameter (for example, a simple collinear magnetic modulation) may induce improper ferroelectricity. The fact that this possibility is not realized at zero field in GdMnO3, for example, results from accidental reasons. Finally, we have also stressed that the symmetry and polar properties of the ordered magnetic phase critically depend on the parity of the modulation wavevector and on the phases of the order parameters.

In addition, the method provides us with well defined predictions about the possible space groups generated by the condensation of a given commensurate order parameter and gives us adequate and symmetry based free energy functionals that have the potential to describe the observed phenomenology.

## APPENDIX

## Complete irreducible co-representations of the paramagnetic group and the magnetic nature of the order parameter

In order to clarify some aspects related to the choice of the basis functions that are relevant to the specification of the displacive or magnetic nature of the order parameter, we will briefly review in this appendix the general method of the calculation of the CICR`s

In the case pertaining to $RMnO_3$, the paramagnetic space group of the reference phase is Pnma and the modulation wavevector of the order parameter is $\vec{k} = (\delta(T),0,0)$. This vector corresponds to the wavevector $k_7$ in Kovalev`s tables [37]. The unitary symmetry elements of the magnetic space group of the reference phase are the following:

$$\{E;000\},\left\{C2x;\frac{1}{2}\frac{1}{2}\frac{1}{2}\right\},\left\{C2y;0\frac{1}{2}0\right\},\left\{C2z;\frac{1}{2}0\frac{1}{2}\right\},\{i;000\},\left\{\sigma x;\frac{1}{2}\frac{1}{2}\frac{1}{2}\right\},\left\{\sigma y;0\frac{1}{2}0\right\},\left\{\sigma z;\frac{1}{2}0\frac{1}{2}\right\}$$

The unitary point group of the wavevector, formed by the sub-set of these symmetry elements that leave $\vec{k}$ invariant up to a Bravais translation, is $C_{2v} \equiv \{E, C_{2x}, \sigma_y, \sigma_z\}$. Consequently, following standard methods, one can construct the small irreducible representations of the little group of the vector $\vec{k}$ . These are given in table 1 (where $\varepsilon = e^{-i\pi\delta}$).

**TABLE 1: small irreducible representation of the little group of the vector $\vec{k}$ .**

| | E | $C_{2x}$ | $\sigma_y$ | $\sigma_z$ |
|---|---|---|---|---|
| $a_1$ | 1 | ε | 1 | ε |
| $b_2$ | 1 | -ε | 1 | -ε |
| $a_2$ | 1 | ε | -1 | -ε |
| $b_1$ | 1 | -ε | -1 | ε |

The magnetic group of the wavevector includes, besides the unitary operations considered above, the non-unitary operations that leave the wavevector invariant up to a reciprocal lattice translation. Here, this group is $D_{2h}(C_{2v}) \equiv \{E, C_{2x}, \sigma_y, \sigma_z, i\theta, \theta\sigma_x, \theta C_{2y}, \theta C_{2z}\}$. The other symmetry elements bring $\vec{k} \to -\vec{k}$. Given that $\vec{k}$ is located inside the Brillouin zone, these two wavevectors are non-equivalent and form the star of the wavevector.

Let $\phi$ be a Bloch function of the standard basis of the irreducible representation of the translational sub-group generated by the vector $\vec{k}$, that is, an eigenfunction of a translation $\{E;\vec{t}\}$ with the eigenvalue $e^{-i\vec{k}\cdot\vec{t}}$

$$\{E;\vec{T}\}\phi = e^{-i\vec{k}\cdot\vec{t}}\,\phi \tag{5}$$

Because of the different behaviour of a displacive or magnetic order parameter, we need to consider how the function $\phi$ is transformed under the action of non-unitary operations. Consider, for example, the anti-unitary operation $a_0 = \theta\{i;000\}$. Through the action of this operation over $\phi$ we will obtain another function $\phi' = a_0\phi$. It can be easily shown that $\phi'$ is also an eigenfunction of $\{E;\vec{t}\}$, sharing with $\phi$ the same eigenvalue [37]:

$$\{E;\vec{t}\}a_0\phi = e^{-i\vec{k}\cdot\vec{t}}a_0\phi \tag{6}$$

These two basis functions, $\phi$ and $a_0\phi$ , will lead to a basis of the small co-representation of the magnetic little group of the vector $\vec{k}$ . Following the standard method [37 49 50], this small co-representation is given by the matrices:

$$D(R) = \begin{bmatrix} \Delta(R) & 0 \\ 0 & \Delta^*(a_0^{-1}Ra_0) \end{bmatrix}$$

(7)

$$D(B) = \begin{bmatrix} 0 & \Delta(Ba_0) \\ \Delta^*(a_0^{-1}B) & 0 \end{bmatrix}$$

Here, $R$ and $B$ denote general unitary and anti-unitary operators, respectively, and $\Delta$ represents the matrix of the corresponding unitary operator in a given irreducible small representation (which in the present case are just complex numbers, see Table 1). Note that $Ba_0$ is unitary. Note also that, given the definition of anti-unitary operator, it follows that [38 51]:

*D(RS)=D(R)D(S); D(RB)=D(R)D(B); D(BR)=D(B)D*$^*$*(R) and D(BC)=D(B)D*$^*$*(C)*

In the case under analysis, $\Delta(R) = \Delta^*(a_0^{-1}Ra_0)$ because $\Delta(a_0^2) = +1$ [38 39]. The set of matrices thus obtained is reducible (small co-representation of the type-a [38]). For example, if we consider the unitary transformation:

$$U=\begin{bmatrix}\frac{1}{\sqrt{2}} & \frac{1}{\sqrt{2}}\\ -\frac{1}{\sqrt{2}} & \frac{1}{\sqrt{2}}\end{bmatrix},$$

we obtain a set of transformed matrices $D'(R)=U^{-1}D(R)U$ and $D'(B)=U^{-1}D(B)U^*$ that are diagonal. The diagonal blocks that correspond to the irreducible co-representations are:

$$d(R)=\Delta(R) \qquad (8a)$$

$$d(B)=\pm\Delta(Ba_0^{-1}) \qquad (8b)$$

The two irreducible small co-representations with the signs ($\pm$) in (8b) correspond to the choice of different base functions given by

$$\psi_+=\frac{1}{\sqrt{2}}(\phi+a_0\phi) \text{ or } \psi_-=\frac{1}{\sqrt{2}}(-\phi+a_0\phi) \qquad (9)$$

Each of these functions behaves differently (symmetrically or anti-symmetrically) under the operation $a_0$ ($a_0\psi_\pm=\pm\psi_\pm$). The choice of $\psi_+$ or $\psi_-$ expresses therefore a choice for a symmetric or anti-symmetric behaviour of the basis of the small co-representation under the action of this anti-unitary operation. Note that the component of a primary order parameter $S_{\vec{k}}$ will be transformed under the action of $a_0$ as $a_0S_{\vec{k}}=\pm S_{\vec{k}}$, depending on its displacive (+) or magnetic (-) nature. If the amplitude of the order parameter is expressed by a real number $S_0$, then this imposes that the standard basis is transformed as $a_0\phi=+\phi$ or $a_0\phi=-\phi$, for a displacive or magnetic order parameter, respectively. In the first case $\psi_+\propto\phi$ and $\psi_-=0$, while in the second case $\psi_+=0$ and $\psi_-\propto\phi$. Note also that, with $j=\sqrt{-1}$, $a_0(j\psi_\pm)=\mp j\psi_\pm$ and therefore $j\psi_\pm\propto\psi_\mp$. We can therefore express a magnetic (displacive) order parameter in the basis $\psi_+$ ($\psi_-$) by ascribing to its vector a purely imaginary amplitude. In this sense, the two bases can be seen as equivalent.

The set of small irreducible co-representations of the magnetic little group, which can be obtained from (8a,b) , is given in Table 2. Here, the upper or lower sign corresponds to the choice of the basis $\psi_+$ or $\psi_-$, respectively and $\varepsilon^*$ denotes the complex conjugate of $\varepsilon$.

**TABLE 2: Small irreducible co-representations of the magnetic group of the vector $\vec{k}$**

| | E | $C_{2x}$ | $\sigma_y$ | $\sigma_z$ | $i\theta$ | $\theta\sigma_x$ | $\theta C_{2y}$ | $\theta C_{2z}$ |
|---|---|---|---|---|---|---|---|---|
| $A_1$ | 1 | ε | 1 | ε | ±1 | ±ε* | ±1 | ±ε* |
| $B_2$ | 1 | -ε | 1 | -ε | ±1 | ∓ε* | ±1 | ∓ε* |
| $A_2$ | 1 | ε | -1 | -ε | ±1 | ±ε* | ∓1 | ∓ε* |
| $B_1$ | 1 | -ε | -1 | ε | ±1 | ∓ε* | ∓1 | ±ε* |

The complete irreducible co-representation of the magnetic space group can now be readily obtained by using the standard techniques described in [37]. In the following, we will adopt the basis $\psi_-$ and, consequently, choose to describe a magnetic order parameter as a vector with a real amplitude and a displacive order parameter as a vector with a purely imaginary amplitude. Following the standard method, and considering the physically equivalent complex conjugated small co-representation related to the vector $-\vec{k}$ of the same star, we readily obtain the CICR matrices that are given in Table 3 for the generators of the magnetic space group ($C_{2x}$, $\sigma_y$, $i$ and $i\theta$). The matrices corresponding to the other symmetry elements can then be obtained from the multiplication table of the group and by taking into account that the anti-unitary operators conjugate the coefficients of the matrices upon which they act.

**TABLE 3: Matrices representing the generators of the group (Pnma)´ in the four of its complete irreducible co-representations at $\vec{k} = \delta \vec{a}_1^*$.**

| | $C_{2x}$ | $\sigma_z$ | iθ | i |
|---|---|---|---|---|
| Γ($A_1$) | $\begin{bmatrix} \varepsilon & 0 \\ 0 & \varepsilon^* \end{bmatrix}$ | $\begin{bmatrix} \varepsilon & 0 \\ 0 & \varepsilon^* \end{bmatrix}$ | $\begin{bmatrix} -1 & 0 \\ 0 & -1 \end{bmatrix}$ | $\begin{bmatrix} 0 & 1 \\ 1 & 0 \end{bmatrix}$ |
| Γ($B_2$) | $\begin{bmatrix} -\varepsilon & 0 \\ 0 & -\varepsilon^* \end{bmatrix}$ | $\begin{bmatrix} -\varepsilon & 0 \\ 0 & -\varepsilon^* \end{bmatrix}$ | $\begin{bmatrix} -1 & 0 \\ 0 & -1 \end{bmatrix}$ | $\begin{bmatrix} 0 & 1 \\ 1 & 0 \end{bmatrix}$ |
| Γ($A_2$) | $\begin{bmatrix} \varepsilon & 0 \\ 0 & \varepsilon^* \end{bmatrix}$ | $\begin{bmatrix} -\varepsilon & 0 \\ 0 & -\varepsilon^* \end{bmatrix}$ | $\begin{bmatrix} -1 & 0 \\ 0 & -1 \end{bmatrix}$ | $\begin{bmatrix} 0 & 1 \\ 1 & 0 \end{bmatrix}$ |
| Γ($B_1$) | $\begin{bmatrix} -\varepsilon & 0 \\ 0 & -\varepsilon^* \end{bmatrix}$ | $\begin{bmatrix} \varepsilon & 0 \\ 0 & \varepsilon^* \end{bmatrix}$ | $\begin{bmatrix} -1 & 0 \\ 0 & -1 \end{bmatrix}$ | $\begin{bmatrix} 0 & 1 \\ 1 & 0 \end{bmatrix}$ |

[6] J. R. Teague, R. Gerson and W. J. James, Solid State Commun. **8** 1073, (1970).

[7] E. F. Bertaut, in *Magnetism*, edited by G. T. Rado, H. Suhl (Academic Press, New York, 1963), Vol. **III**, 149.

[8] T. Kimura, S. Kawamoto, I. Yamada, M. Azuma, M. Takano and Y. Tokura, Phys. Rev. B **67**, 180401 (2003).

[9] Landolt-Börnstein, *Numerical Data and Functional Relationships, new Series, Group III*, (Springer-Verlag, Berlin, 1981), Vol. **16a**.

[10] Z. J. Huang, Y. Cao, Y. Y. Sun, Y. Y. Xue and C. W. Chu Phys. Rev. B **56**, 2623 (1997).

[11] M. N. Iliev, H. G. Lee, V. N. Popov, M. V. Abrashev, A. Hamed, R. L. Meng and C. W. Chu, Phys. Rev B **56**, 2488 (1997).

[12] T. Kimura, T. Goto, H. Shintani, K. Ishizaka, T. Arima and Y. Tokura, Nature **426**, 55 (2003).

[13] N. Hur, S. Park, P. A. Sharma, J. S. Ahn, S. Guha and S. W. Cheong, Nature **429**, 392 (2004).

[14] G. Lawes, A. B. Harris, T. Kimura, N. Rogado, R. J. Cava, A. Aharony, O. Entin-Wohlman, T. Yildirim, M. Kenzelmann, C. Broholm and A. P. Ramirez, Phys. Rev. Lett. **95**, 087205 (2005).

[15] T Kimura, J.C. Lashley, and A. P. Ramirez, Phys. Rev. B **73**, 220401(R) (2006).

[16] Y Yamasaki, Phys. Rev. Lett. **96**, 207204 (2006).

[17] K. Taniguchi, N. Abe, T. Takenobu, Y. Iwasa, and T. Arima, Phys. Rev. Lett. **97**, 097203 (2006).

[44] C. J. Bradley, J. of Math. Phys. **7**, 1145 (1966).

[45] M. Lax and J. J. Hopfield, Phys. Rev. **124**, 115 (1961).

[46] M. Lax, Phys. Rev. **3A**, A793 (1965).

[47] I. Dzyaloshinskii, Sov. Phys. JETP **19**, 960 (1964).

[48] T. Moriya, Phys. Rev. **120**, 91 (1960).

[49] C. J. Bradley and B. L. Davies, Rev. Mod. Phys. **40**, 359 (1968).

[50] E. P. Wigner, *Group Theory and its Applications to Quantum Mechanics of Atomic Spectra* (Academic Press, 1960)

[51] E. P. Wigner, J. of Math. Phys. **1**, 409 (1960).